\documentclass[aps,prl,twocolumn,groupedaddress]{revtex4} 
\usepackage{graphicx}
\usepackage{bm}

\begin{document}

\title{Bose-Einstein Condensation of $^{84}$Sr}
\author{Y. N. Martinez de Escobar, P. G. Mickelson, M. Yan, B. J. DeSalvo, S. B. Nagel and T. C. Killian}
\affiliation{Rice University, Department of Physics and
Astronomy, Houston, Texas, 77251}

\date{\today}

\begin{abstract}

We report Bose-Einstein condensation of $^{84}$Sr in an optical dipole trap. Efficient laser cooling on the
narrow intercombination line and an ideal s-wave scattering length  allow creation of large condensates $(N_0\sim
3 \times 10^5)$ even though the natural abundance of this isotope is only 0.6\%. Condensation is heralded by the
emergence of a low-velocity component in time-of-flight images.

\end{abstract}


\maketitle

The study of quantum degenerate gases continues to be at the
forefront of research in atomic and condensed matter physics nearly
15 years after the first observation of Bose-Einstein condensation
\cite{aem95,dma95,bst95}. Current areas of focus include the
behavior of quantum fluids in optical lattices \cite{gfo08}, effects
of dimensionality \cite{gvl01} and disorder \cite{bjz08, ref08},
exploration of the BEC-BCS crossover regime \cite{bdz08}, and the
pursuit of quantum degenerate molecular systems \cite{nom08}.

Quantum degeneracy in alkaline earth metal atoms and atoms with similar
electron structure has become another area of intense activity. These systems
have been the subject of many recent theoretical proposals for quantum
computing in optical lattices \cite{dby08,grd09, rjd09} and creation of novel
quantum fluids \cite{hgm09}. They possess  interesting and useful collisional
properties, such as a wealth of isotopes that allow mass-tuning of interactions
\cite{kek08,mmp08} and creation of various quantum mixtures \cite{fst09}.
Low-loss optical Feshbach resonances  \cite{ctj05,ekk08,mmy09} in these atoms
 promise new opportunities because they enable changing the atomic
 scattering lengths on small spatial and
temporal scales. This can lead to creation of matter-wave solitons in two dimensions \cite{sue03} and random
nonlinear interactions in quantum fluids \cite{fwg89}. Many of these ideas take advantage of the existence of
long-lived metastable triplet states and associated narrow optical transitions, which are also the basis for
recent spectacular advances in optical frequency metrology \cite{ykk08}.

Here we report the observation of Bose-Einstein condensation of
$^{84}$Sr. To date, quantum degeneracy with alkaline earth metal
atoms and similar elements has been reported in a collection of
ytterbium isotopes \cite{fst09,fst07,ftk07,tmk03}, and very recently
in $^{40}$Ca \cite{kva09} and $^{84}$Sr \cite{sth09}, demonstrating
increasing interest in these systems. Starting with a
natural-abundance strontium source, our success in forming a
relatively large Bose-Einstein condensate of $^{84}$Sr, which has a
natural abundance of 0.6\%, demonstrates the power of laser cooling
in strontium using the $(5s^2)^1$S$_0$-$(5s5p)^3$P$_1$ narrow
intercombination line at 689 nm \cite{kii99}. It also reflects the
near-ideal scattering properties of this isotope, which has an
$s$-wave elastic scattering length of $a=122.7(3)$\,$a_0$
\cite{mmp08}, where $a_0=0.53$ \AA. The ground state scattering
properties of all Sr isotopes are well-characterized from one-
\cite{nms05,mms05,ykt06} and two-photon \cite{mmp08}
photoassociative spectroscopy, fourier transform molecular
spectroscopy \cite{skt08}, and experiments trapping various isotopes
in optical traps \cite{fdp06,pdf05}.

\begin{figure}
\includegraphics[clip=true,keepaspectratio=false,width=3.5in,height=2.5in]{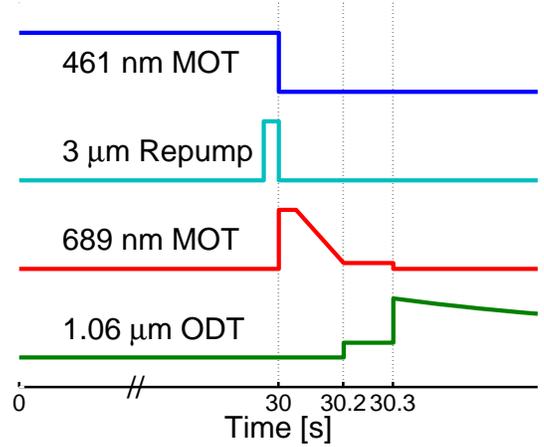}\\
\caption{(color online) Timing diagram for principal experimental components.
The traces are offset for clarity and heights schematically indicate variation
in power. Note the break in the time axis. }
 \label{Timing Diagram}
\end{figure}

Our apparatus and experimental sequence for laser cooling atoms and
loading them into our optical dipole trap were described in
\cite{nsl03,mmp08,mma09}, and a timing diagram is shown in Fig.\
\ref{Timing Diagram}. Several techniques are used to enhance initial
number and subsequent transfers to overcome the low abundance of
$^{84}$Sr. We form an atomic beam and utilize 2-dimensional (2D)
collimation and Zeeman cooling to load a magneto-optical trap (MOT),
all using the $(5s^2)^1$S$_0$-$(5s5p)^1$P$_1$ transition at 461 nm.
The MOT consists of three retro-reflected beams, each with peak
intensity of 3.6 mW/cm$^2$ and red-detuned from atomic resonance by
60\,MHz.

One in $10^5$  $(5s5p)^1$P$_1$ atoms decays to the $(5s4d)^1$D$_2$ level and then to the $(5s5p)^3$P$_1$ and
$(5s5p)^3$P$_2$ states. $(5s5p)^3$P$_1$ atoms return to the ground state and are recaptured in the MOT, but a
fraction of $(5s5p)^3$P$_2$ atoms, with a lifetime of 9\,min \cite{yka04}, are trapped in the quadrupole magnetic
field of the MOT \cite{lbm02,nsl03,mma09}. We can accumulate $2.5 \times 10^7$ atoms in an axial magnetic field
gradient of 60 G/cm with 30 s of loading. $(5s5p)^3$P$_2$ atoms are then returned to the 461\,nm cycling
transition by repumping for 35\,ms via the $(5s5p)^3$P$_2$-$(5s4d)^3$D$_2$ transition at 3012 nm \cite{mma09}. We
subsequently reduce the 461\,nm beam intensities to about 0.36 mW/cm$^2$ for 6.5 ms to reduce the sample
temperature to 2\,mK.

 The 461 nm light is then extinguished, and a second stage of
cooling begins using the 689 nm $(5s^2)^1$S$_0$-$(5s5p)^3$P$_1$
intercombination line. This MOT light, consisting of three
retro-reflected beams with 2 cm diameters, is initially detuned 1
MHz red of resonance and broadened by 700 kHz (peak-to-peak dither
amplitude) to enhance our 689 nm MOT capture rate. The peak
intensity is 0.75 mW/cm$^2$ per beam, and the magnetic quadrupole
field gradient is 0.1 G/cm. Over the next 150\,ms, the field
gradient is increased to 0.8 G/cm, the laser spectrum is reduced to
single frequency and detuned by only $\sim 30$\,kHz, and the power
is reduced to 0.15\,mW/cm$^2$. This results in 1.6 $\times$ $10^7$
atoms at a temperature of  1 $\mu$K and a peak density of $\sim
10^{12}$\,cm$^{-3}$.

 An
optical dipole trap (ODT) consisting of two crossed beams is then overlapped with the intercombination-line MOT
for 115 ms with modest power (2.5 W) per beam. The ODT is formed by a single beam derived from a 20 W multimode,
1.06\,$\mu$m fiber laser that is recycled through the chamber. No attempt is made to change the polarization of
the beam during recycling. The beams cross nearly perpendicularly, but the plane of the lasers is inclined by
approximately $10.5^\circ$ from horizontal. This results in a trap with equipotentials that are nearly oblate
spheroids, with the tight axis close to vertical. Each beam has a waist of approximately 100 $\mu$m in the
trapping region. This is slightly different than other realizations of quantum degeneracy in two-electron atoms,
in which at least one beam had a waist that was significantly smaller \cite{fst09,ftk07,fst07, kva09,sth09}.

Immediately after extinction of the 689 nm light, the ODT  power is ramped in 20 ms to 10 W to obtain a sample of
$3 \times 10^6$ atoms at 5 $\mu$K. The trap depth is 31 $\mu$K, and the peak density at this point is $4 \times
10^{13}$ cm$^{-3}$, which implies an average collision rate of 1000\,s$^{-1}$. The peak phase space density (PSD)
is $10^{-2}$. For diagnostics, we record absorption images of samples released from the trap after a time of
flight varying from 10 to 40 \,ms, and the optical depth profile can be related to the areal density, which, for
these long delays, also provides the velocity distribution.

\begin{figure}
\centering
\includegraphics[clip=true,keepaspectratio=true,width=3in]{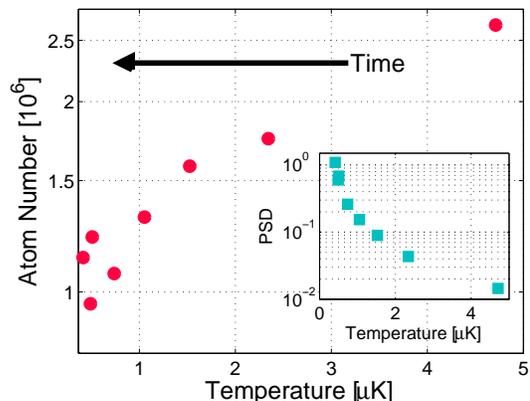}\\
\vspace{-.2in} \caption{(color online) Number, temperature and phase space
density along a typical evaporation trajectory, from 200\,ms after the
completion of loading of the optical dipole trap and commencement of
evaporation, until quantum degeneracy is reached 3\,s later. The phase space
density increases by a factor of 100 for a reduction of atom number by a factor
of 3. } \label{Evaporation Trajectory}
\end{figure}

Figure \ref{Evaporation Trajectory} shows the number, temperature, and phase
space density evolution for  a typical forced evaporation trajectory. We ramp
down the power in the lasers according to $P=P_0/(1+t/\tau)^{\beta}$, with time
denoted by $t$,  $\beta=1.5$, and $\tau=1.5$\,s. This trajectory is designed to
approximately maintain a constant ratio of trap depth to temperature ($\eta$)
during evaporation \cite{ogg01}. The lifetime of atoms in the ODT is 30 s,
limited primarily by background gas collisions. This allows efficient
evaporation at a measured $\eta= 8 \pm 1$   for most of the trajectory.

After 3 s of evaporation to a power of 2\,W, $1\times10^6$ atoms
remain at a temperature of 0.4 $\mu$K. For this trap, our measured
trap oscillation frequencies \cite{fdw98} yield a trap depth,
including the effect of gravity, of $3.5 \pm 0.3$\,$\mu$K and a mean
trap frequency of $\bar{f}=(f_x f_y f_z)^{1/3}=84\pm 5$ Hz. These
values indicate that the sample is at the
 critical transition temperature for a harmonic trap \cite{psm02},
\begin{equation}\label{Equation: Transition Temperature}
T_c=\frac{1}{k_B}\frac{\hbar \overline{\omega} N^{1/3}}{\zeta(3)},
\end{equation}
at which the PSD is 1.2, where $N$ is the number of trapped atoms,
$k_B$ is Boltzmann's constant, $\hbar$ is Planck's constant,
$\overline{\omega}$ is 2$\pi \bar{f}$, and $\zeta$ is the Riemann
zeta function.

The evaporation is quite efficient. We lose a factor of 3 in the number of atoms from initiation of forced
evaporation to the onset of degeneracy, while the phase space density increases by about a factor of 100. The
predicted scaling $\rho/\rho_i=(N_i/N)^{\eta'-4}$ \cite{ogg01}, where $\rho_i$ and $N_i$ are the initial phase
space density and number, and $\eta'= \eta + (\eta - 5)/(\eta - 4)$, implies that $\eta=7.5$, in excellent
agreement with our estimate from knowledge of the optical trap parameters and measured sample temperature.

\begin{figure}
\includegraphics[clip=true,keepaspectratio=true,width=3.5in, trim=50 0 0 0 ]{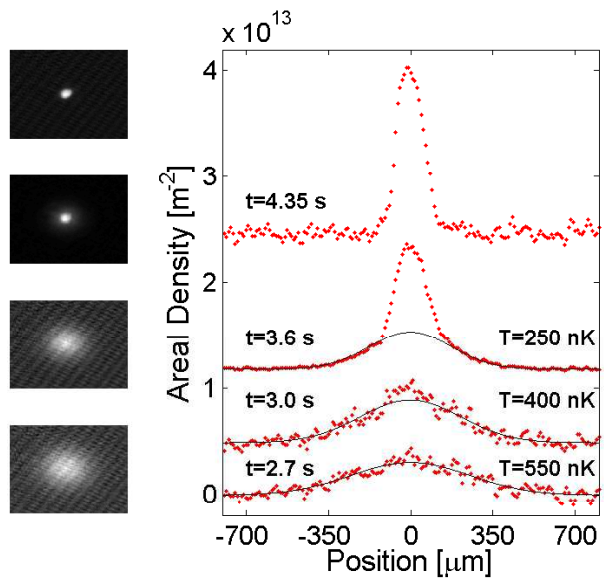}\\
\caption{(color online) Appearance of Bose-Einstein condensation in absorption images (left) and areal density
profiles (right). Data corresponds to 35 ms of free expansion after indicated evaporation times (t). (left)
Images are 1.8 mm per side, and have the same time stamp  as the density profiles. (right) The areal density
profiles are from a vertical cut through the center of the atom cloud, and temperatures are extracted from 2D
gaussian fits to the thermal component. At 3.0 and 3.6 s, a bimodal distribution indicative of Bose-Einstein
condensation becomes increasingly clear, and then  a pure condensate
 is shown at 4.35 s.
} \label{Bimodal Distribution}
\end{figure}

Figure \ref{Bimodal Distribution} shows false color 2D rendering
 and
1D slices through the time-of-flight images recorded after 38\,ms of
expansion for various points along the evaporation trajectory. At
2.7 s of evaporation, the distribution is fit well by a Boltzmann
distribution, but at 3\,s, the presence of a Bose-Einstein
condensate is indicated by the emergence of a narrow peak at low
velocity. Further evaporation to a beam power of 1.3 W at 4.35\,s
and trap depth of 600\,nK yields a condensate with negligible
discernible thermal fraction.

The areal density of expanded condensates is fit with the functional form
\cite{dgp99}
\begin{equation}\label{Equation: Thomas-Fermi expansion}
    n(x,y)=\frac{5 N_0}{2\pi}\left(1-\frac{x^2}{R_x^2}-\frac{y^2}{R_y^2}\right)^{3/2}
    \theta\left(1-\frac{x^2}{R_x^2}-\frac{y^2}{R_y^2}\right),
\end{equation}
 where
$\theta$ is the Heaviside function, and $R_x$ and $R_y$ are the condensate radii. At 600\,nK trap depth, fitting
the areal density yields typical values of $N_0= 3.0\times 10^5$ for the largest condensates. The trap mean
oscillation frequency at this point is $\bar{f}=70$ Hz and harmonic oscillator length $a_{ho}=(\hbar/(M
\overline{\omega}))^{1/2} = 1.3$ $\mu$m. The trap harmonic oscillator energy scale is $\hbar \overline{\omega}/k_
B=3.3$ nK and the chemical potential, $\mu = \frac{\hbar \overline{\omega}}{2 k_B} (\frac{15 N
a}{a_{ho}})^{2/5}=90$ nK. $(N_0 a)/a_{ho}=1500$ is large, implying that the condensate is in the strong
interaction regime and should be well described by a Thomas-Fermi density distribution, confirming that Eq.\
\ref{Equation: Thomas-Fermi expansion} is the appropriate description.

\begin{figure}
\includegraphics[clip=true,keepaspectratio=true,width=3in,height=3in]{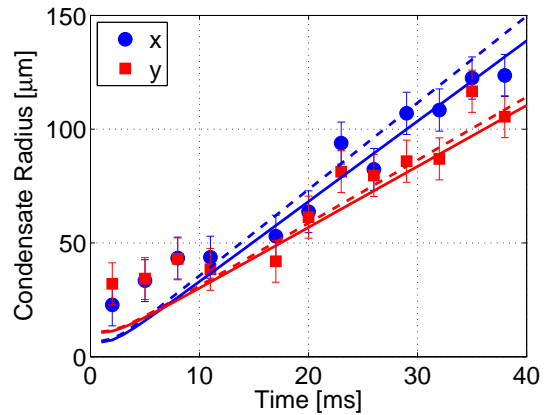}\\
\caption{(color online) Condensate radius after release from the
trap. The expansion of the condensate is fit by a model described in
the text. Deviations at early times most likely reflect limitations
of the optical imaging system and large optical depths of the
condensate. } \label{Condensate Radius}
\end{figure}

By releasing a pure condensate from the trap and varying the time of
flight before recording an absorption image, we are able to
characterize the expansion of the many-body wavefunction which is
sensitive to the chemical potential and confinement of the trapped
sample. Figure \ref{Condensate Radius} shows the evolution of
condensate radii as a function of expansion time obtained by fitting
the optical depth to Eq.\ \ref{Equation: Thomas-Fermi expansion}.
The fitting coordinate system has been rotated to align with the
principle axes of the condensate. These samples contain
$N_0=2.3\times 10^5$ atoms and were obtained after 4.5\,s of
evaporation to an estimated trap depth of 350\,nK and trap
frequencies of 90\,Hz close to vertical and 60\,Hz along the
perpendicular axes. The radii show an inversion, as expected for an
expanding condensate, but the effect is small because the trap is
not highly asymmetric.

We fit the data using a numerical solution of the Castin-Dum model \cite{cdu96} for the free expansion of a
Thomas-Fermi wavefunction \cite{dgp99}. The radii evolve according to
\begin{equation}\label{Equation: castin dum}
   R_i(t)= R_i(0)b_i(t)= \sqrt{\frac{2\mu}{m(2\pi f_i)^2}}b_i(t)
\end{equation}
for $i=x,y,z$, where the scaling parameters obey
$\ddot{b}_i(t)=(2\pi f_i)^2/(b_ib_xb_yb_z)$. Using the well-known
value of the scattering length \cite{mmp08} and our estimated trap
parameters obtained from beam profiling and trap oscillation
measurements, we obtain the dashed line in Fig.\ \ref{Condensate
Radius}. Decreasing the frequencies by about 5\% yields an improved
fit (solid line), which is within our uncertainty in trap
parameters.

\begin{figure}

\includegraphics[clip=true,keepaspectratio=true,width=3in,height=3in]
{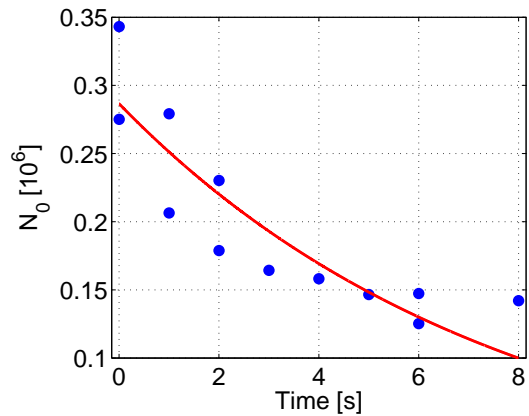}\\
\caption{(color online) Number of condensate atoms (filled circles) held in the optical dipole trap at constant
depth versus time after the evaporation (see text). The solid line is an exponential fit of the data, yielding a
condensate lifetime of 8$\pm$1 s.} \label{BEC Lifetime}
\end{figure}

If the trap is held constant after 4.5\,s of evaporation, instead of releasing
the condensate from the trap, the lifetime of the condensate can be measured.
Figure \ref{BEC Lifetime} shows a fit of the decay to an exponential, yielding
a lifetime of $8\pm 1$\,s. No discernable thermal fraction was evident in these
samples.

The large number of condensate atoms and long lifetime indicate favorable conditions for many experiments. Of
particular interest in the near future are the achievement of quantum degeneracy with other strontium isotopes
and mixtures and the application of an optical Feshbach resonance to a quantum degenerate sample to modify the
scattering length on small spatial and temporal scales.

\textmd{\textbf{Acknowledgements}}

We thank F. Schreck for helpful discussions and F. Tittel for the loan of
crucial equipment. This research was supported by the Welch Foundation
(C-1579), National Science Foundation (PHY-0855642), and the Keck Foundation.


\end{document}